\documentclass[10pt, technote]{IEEEtran}
\usepackage{cite}
\usepackage[cmex10]{amsmath}
\usepackage[dvips]{graphicx}
\usepackage{subfigure,color}
\usepackage{float}
\usepackage{subeqnarray}
\usepackage{cases}

\usepackage{epsf}
\usepackage{amsfonts}
\usepackage{amsmath}
\usepackage{graphics,graphicx}
\usepackage{subfigure}
\usepackage{bm}
\usepackage{epsfig}
\usepackage{placeins}
\usepackage{multirow}
\usepackage{times}
\usepackage{amsmath,bm}
\usepackage{color}

\usepackage{color}

\begin{document}

\title{Ultrathin Complementary Metasurface for Orbital Angular Momentum Generation at Microwave Frequencies}

\author{Menglin~L.~N.~Chen,
        Li~Jun~Jiang,
        and~Wei~E.~I.~Sha
\thanks{All the authors are with Department of Electrical and Electronic Engineering, The University of Hong Kong, Pokfulam Road, Hong Kong. e-mail: menglin@connect.hku.hk (M. L. N. Chen), ljiang@eee.hku.hk (L. J. Jiang) and wsha@eee.hku.hk (W. E. I. Sha).}}


\maketitle

\begin{abstract}
Electromagnetic (EM) waves with helical wavefront carry orbital angular momentum (OAM), which is associated with the azimuthal phase of the complex electric field. OAM is a new degree of freedom in EM waves and is promising for channel multiplexing in communication system. Although the OAM-carrying EM wave attracts more and more attention, the method of OAM generation at microwave frequencies still faces challenges, such as efficiency and simulation time. In this work, by using the circuit theory and equivalence principle, we build two simplified models, one for a single scatter and one for the whole metasurface to predict their EM responses. Both of the models significantly simplify the design procedure and reduce the simulation time. In this paper, we propose an ultrathin complementary metasurface that converts a left-handed (right-handed) circularly polarized plane wave without OAM to a right-handed (left-handed) circularly polarized wave with OAM of arbitrary orders and a high transmission efficiency can be achieved.
\end{abstract}

\begin{IEEEkeywords}
Orbital angular momentum, ultrathin complementary metasurface, circuit theory, equivalence principle, Babinet's principle.
\end{IEEEkeywords}

\section{Introduction}
\IEEEPARstart{I}{n} 1992, Allen first experimentally demonstrated that a light beam with helical wavefront carries orbital angular momentum (OAM)~\cite{allen1992orbital}. Unlike the spin angular momentum (SAM) of photon that only takes values of $\pm\hbar$ and $0$, OAM takes unlimited number of values of $l\hbar$, where $l$ is any integer and $\hbar$ is the Planck constant. Each OAM value corresponds to an eigenmode of electromagnetic (EM) wave, which can be employed in the multiplexing and de-multiplexing in communication channels~\cite{wang2012terabit,padgett,Nanoscale}. On the other hand, technical challenges exist in practical applications of OAM-carrying wave~\cite{real}. For example, the divergence nature of the OAM-carrying wave causes a practical issue in the OAM detection process: large receiving aperture is needed to collect high enough power to guarantee the performance of the system. Additionally, to effectively detect the phase profile of the OAM-carrying wave, the alignment between the transmitter and receiver is necessary, or else the orthogonality of different OAM modes will be broken, resulting in the crosstalk between adjacent OAM modes. Meanwhile, the OAM-carrying wave is vulnerable to the turbulence in the atmosphere which may make it unsuitable for long-distance communication~\cite{aberration, decoherence}. Efforts have been made to improve the performance of OAM based multiple-input multiple-output (MIMO) communication system. In~\cite{misalign}, parameters that affect the OAM based MIMO link have been examined with a simulation model. The controversy about the performance of OAM based and the conventional MIMO systems still exists~\cite{zhang_mimo,smile}.

To generate OAM-carrying EM waves, we need to introduce an azimuthal phase dependence $e^{il\phi}$ into EM waves, where $l$ is the OAM index and $\phi$ is the azimuthal angle. There are several approaches to produce $e^{il\phi}$. At optical frequencies, a common method is by modulating the length of optical path, such as the planar spiral phase plate (SPP)~\cite{SPP_SREP, zhang_awpl} and logarithmic-spiral zone plates~\cite{aom_qiu}, or introducing abrupt phase shift in the light path using scatterers, such as the metasurface composed of V-shaped scatterers~\cite{CapassoScience}. At microwave frequencies, a loop antenna with the presence of $e^{il\phi}$ in its surface current to radiate OAM-carrying wave is proposed~\cite{zhangxianmin}. In~\cite{patch}, a patch antenna was well designed to radiate circularly polarized wave with different orders of OAM. However, excitation of high-order modes needs to be carefully implemented so that a pure OAM mode can be obtained. Another approach to introduce  $e^{il\phi}$ is by utilizing spin-to-orbital coupling effect~\cite{APL_spin_to_orbital,LSA_spin_to_orbital}. According to the momentum conservation law and Pancharatnam-Berry phase concept~\cite{menglin}, the SAM could be converted to OAM. This effect requires a retardation of $\pi$ between two orthogonal linear polarizations and the incidence of circularly polarized wave. However, the retardation of $\pi$ and a high transmission (conversion) efficiency are difficult to achieve simultaneously. At microwave frequencies, the universal challenge also arises. Researchers made use of multilayer structures to gain this $\pi$ phase difference and meanwhile to preserve the transmission efficiency. Several prototypes have been proposed~\cite{motl,AWPL}. However, the designs are very complicated and hard to fabricate concerning the unit-cell size, substrate thickness and number of printed circuit board (PCB) layers. On the other hand, the whole design process is time consuming due to the complicated fine structure and large simulation domain.

All these deficiencies motivate us to improve the design and also, to develop a novel modelling methodology which could facilitate a fast and simple design process. In this work, we propose an ultrathin complementary metasurface composed of sub-wavelength scatterers. It transmits a circularly polarized plane wave to a OAM-carrying wave of desired order with a high efficiency. Section II describes the basic working principle. Section III discusses the design of sub-wavelength scatterers by adopting an equivalent circuit model. The whole metasurface is built and simulated in Section IV. Firstly, according to equivalence principle, the sub-wavelength scatterers are replaced by equivalent magnetic dipole sources. Then we calculate the EM response with the help of magnetic dyadic Green's function. Afterward, full-wave simulation results are provided to validate the simplified model. At last, the efficiency of the metasurface is examined.

\section{Theory}
We consider a monochromatic plane wave propagating along the $z$ direction. The incident field $\mathbf{E}_i$ is decomposed into two orthogonal components, $E_{ix}$ and $E_{iy}$. When the wave impinges on a metasurface, the local behavior of a scatterer on the metasurface is described by a transmission matrix $\mathbf{T}$. It connects the transmitted field components, $E_{tx}$ and $E_{ty}$ to the incident ones,

\begin{equation}
\begin{pmatrix}
  E_{tx} \\ E_{ty}
\end{pmatrix}
=
\mathbf{T}
\begin{pmatrix}
  E_{ix} \\ E_{iy}
\end{pmatrix}
=
\begin{pmatrix}
    T_{xx} & T_{xy} \\
    T_{yx} & T_{yy}
  \end{pmatrix}
\begin{pmatrix}
  E_{ix} \\ E_{iy}
\end{pmatrix}
\end{equation}

As demonstrated in~\cite{APL_spin_to_orbital}, if $T_{yy}=-T_{xx}$ and $T_{yx}=T_{xy}=0$, by axially rotating the scatterer at an angle of $\alpha$, a new transmission matrix can be derived,

\begin{subequations}
\begin{align}
\mathbf{T}_{l}(\alpha)
&=T_{xx}
\begin{pmatrix}
    \cos(2\alpha) & \sin(2\alpha) \\
    \sin(2\alpha) & -\cos(2\alpha)
  \end{pmatrix},
  \label{4}
\\
\mathbf{T}_{c}(\alpha)
&=T_{xx}
\begin{pmatrix}
  0 & e^{-2i\alpha} \\ e^{2i\alpha} & 0
\end{pmatrix}
\end{align}
\end{subequations}
where the subscripts, $l$ and $c$ stand for the linear and circular basis, respectively.

It should be noted that under the circular basis, an additional phase factor $e^{\pm 2i\alpha}$ that has a similar form to the azimuthal phase dependence $e^{\pm il\phi}$ can be introduced. Under the incidence of linear or dual-linear polarization, according to Eq.~\ref{4}, no such phase factor can be introduced. Therefore, manipulation under the circular basis is necessary. With only the non-zero off-diagonal items in $\mathbf{T}_{c}(\alpha)$, the axially rotated scatterer switches the polarization state from left-handed (right-handed) circular polarization to right-handed (left-handed) circular polarization. By fixing the value of $l$, if we arrange the scatterers in a manner satisfying $\alpha = \l\phi/2$, where $\phi$ is the azimuthal location of each scatterer, OAM of order $\l$ can be introduced in the transmitted EM wave. The value of $\alpha/\phi$ is known as the topology charge $q$ of the metasurface composed of these scatterers. The OAM order $\l$ is determined by $q$ and should be double of its value. The conversion efficiency is determined by the magnitude of $T_{xx}$ or $T_{yy}$. Ideally, one can obtain a perfect (100\%) conversion if $T_{xx}$ and $T_{yy}$ have the same magnitude of $1$ and $\pi$ phase difference. Unfortunately, when considering the reflection at air-dielectric interfaces, dielectric loss, and metal loss, the co-transmission coefficients cannot reach 1.

\section{Scatterer Design}

Based on our analysis, we need to design an anisotropic scatterer satisfying $T_{yy}=-T_{xx}$ and $T_{yx}=T_{xy}=0$. Meanwhile, a high magnitude of $T_{xx}$ or $T_{yy}$ is preferred to guarantee the high conversion efficiency. It is well known that a half-wavelength dipole resonator has a high reflectance at its resonance. According to Babinet's principle, to achieve a high transmittance, we start from the complementary frequency selective surface (FSS)~\cite{munk2005frequency}. Here, we choose the square ring-shaped scatterer so that the size of unit cell is in sub-wavelength scale at its half-wavelength resonance. To realize the anisotropic response, the geometry of the square ring has to be modified to break the four-fold rotational symmetry. Moreover, to fulfill $T_{yx}=T_{xy}=0$, the mirror symmetry with respect to the $xz$ or $yz$ plane needs to be maintained~\cite{PRA_Jones_Calculus}.

\begin{figure}[h]
\centering
\includegraphics[width=0.9\columnwidth]{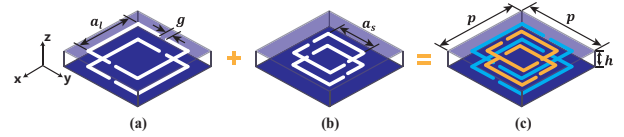}
\caption{Schematic of the proposed scatterer. (a) one bi-layer CSRRs with four U-shaped slots, symmetric along the $x$ direction; (b) the other bi-layer CSRRs with four U-shaped slots, symmetric along the $y$ direction; (c) combined CSRRs. Periodicity is $p$ along both the $x$ and $y$ directions. The distance between the two layers is $h$.}
\label{final_unit}
\end{figure}

Based on above discussions, we propose a complementary FSS as shown in Fig.~\ref{final_unit}. It consists of two types of complementary split-ring resonators (CSRRs) with different orientations, one of which is symmetric along the $x$ direction (Fig.~\ref{final_unit}(a)) and the other along the $y$ direction (Fig.~\ref{final_unit}(b)). Each type of the CSRRs has two pairs of U-shaped slots, which are deposited on the top and bottom layers of a PCB, respectively. The size difference of the slots with the two orientations makes the proposed scatterer anisotropic. Also, the cross-transmission coefficients $T_{yx}$ and $T_{xy}$ are zero due to the mirror symmetry about the $xz$ and $yz$ planes. For a conducting U-shaped strip whose complementary counterpart is same as the structure in Fig.~\ref{final_unit}(a), its fundamental resonance is excited under the $x$ polarized incident wave. According to Babinet's principle, we should exchange the electric field with the magnetic field for the complementary structure, i.e. an $x$ polarized magnetic field is needed to excite the fundamental resonance of the scatterer in Fig.~\ref{final_unit}(a). Consequently, the CSRRs can be excited under the $y$ polarized incident wave and the length of the U-shaped slot is around half of the wavelength at resonance. The fundamental resonance of the other CSRRs in Fig.~\ref{final_unit}(b) can be excited under the $x$ polarized incident wave and the resonant frequency is higher because of their smaller size.

\subsection{Equivalent circuit model}

\begin{figure}[h]
\centering
\includegraphics[width=0.9\columnwidth]{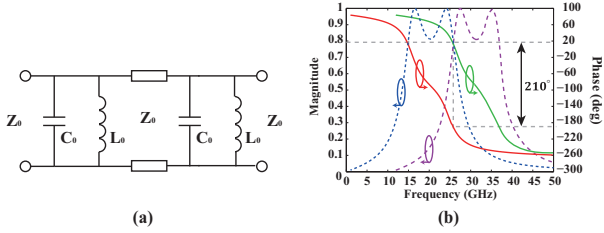}
\caption{Equivalent circuit model of the scatterer in Fig.~\ref{final_unit}(a) and (b). (a) equivalent circuit; (b) simulated $S_{21}$. The purple and green curves are obtained by the translation of the original blue (magnitude) and red (phase) curves. The distance between the two layers is $h$ = 1.25 mm. Characteristic impedance of free space is $Z_0=377~\Omega$. The full-wave simulation results for curve fitting is obtained by simulating the scatterer in Fig.~\ref{final_unit}(a) with the geometric parameters of $a_l=0.5$~mm, $g=2$~mm. The resultant capacitance and inductance are $C_0=0.09$~pF and $L_0=1.03$~nH.}
\label{equivalent}
\end{figure}

The reason why we use the two-layer CSRRs is to achieve a high transmittance as well as the $\pi$ phase difference between the co-transmission coefficients. The two types of CSRRs share the same equivalent circuit model. Therefore, we consider the two types of CSRRs separately. Fig.~\ref{equivalent}(a) illustrates the equivalent circuit model for one type of the CSRRs. They operate like a band-pass filter, so we represent the CSRR at the top (bottom) layer as a parallel inductor-capacitor (LC) circuit; and the space in between is modelled by a transmission delay line. The values of $L_0$ and $C_0$ are obtained by fitting the circuit response in reference to the full-wave simulation results of the CSRR at one layer using CST MWS. For simplicity, the characteristic impedance of the delay line is set to be the same as that of free space. The length of it is set according to the height value $h$. Fig.~\ref{equivalent}(b) shows the simulated transmission coefficients based on the circuit model. At the frequency band of interest, the phase changes from $90^\circ$ to $-260^\circ$, which is much larger than the $180^\circ$ phase change for a single resonator. In other words, the two LC circuits offer a sufficient phase coverage for design optimization.

Fig.~\ref{equivalent}(b) presents useful information on how to realize the $180^\circ$ phase difference. For example, if the resonant frequency is shifted to a higher value (that can be realized by the other type of CSRRs with smaller size), we can make $|\text{phase}(T_{xx})-\text{phase}(T_{yy})|=210^\circ$ and $\text{mag}(T_{xx})=\text{mag}(T_{yy})=0.8$. An exact $180^\circ$ phase difference will be obtained if we decrease the difference of the two resonant frequencies. The smaller frequency difference will result in a higher value of $\text{mag}(T_{xx})$ at the frequency when $\text{mag}(T_{xx})=\text{mag}(T_{yy})$, indicating a high conversion efficiency that is larger than $64\%$.

\subsection{Simulation}

\begin{figure}[h]
\centering
\includegraphics[width=0.9\columnwidth]{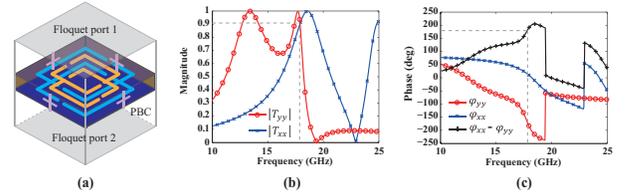}
\caption{Full-wave simulation of the proposed scatterer. (a) schematic and simulation settings; (b) simulated magnitudes of the co-transmission coefficients; (c) simulated phases of the co-transmission coefficients. The period of the unit cell is $7\times7$~mm$^2$. Side lengths of the two types of square CSRRs are $a_l=5.2$~mm and $a_s=3.9$~mm. The length of the gap is $g=0.2$~mm. The width of the slots is $t=0.2$~mm.}
\label{result}
\end{figure}

Full-wave simulation and optimization are done by CST MWS. Two floquet ports and two sets of periodic boundaries are imposed at the longitudinal sides and four lateral sides, respectively, as illustrated in Fig.~\ref{result}(a). F4B220 is chosen as the dielectric substrate. The relative permittivity and thickness are $\epsilon_{r}=2.2$ and $h=0.8$~mm. The loss tangent is $0.003$.

As illustrated in Fig.~\ref{result}(b) and (c), at $17.85$~GHz, $\text{mag}(T_{xx})=\text{mag}(T_{yy})=0.91$ and $|\text{phase}(T_{xx})-\text{phase}(T_{yy})|=180^\circ$, indicating a $81\%$ right-to-left (left-to-right) circular polarization conversion efficiency. The shape of the magnitude curve by the full-wave simulation is slightly different from that in Fig.~\ref{equivalent}(b) by the circuit model. Because in the circuit model, we only consider the fundamental resonance and each scatterer is assumed to only response to the $x$ or $y$ polarized wave. However, in real case, high-order resonant modes exist. Except for the fundamental half-wavelength mode, the most significant mode is the high-order full-wavelength resonant mode. The full-wavelength resonant mode of the large CSRRs (Fig.~\ref{final_unit}(a)) is excited by $x$ polarized wave and interacts with the fundamental resonant mode of the small CSSRs (Fig.~\ref{final_unit}(b)). Considering the length of CSRRs, the resonant wavelength has the following relationship: $\lambda_{0y}^{l}>\lambda_{0x}^{s}>\lambda_{1x}^{l}>\lambda_{1y}^{s}$, where $x$ and $y$ denote the polarization state of incident wave, $l$ and $s$ denote large and small CSRRs, and $0$ and $1$ denote fundamental half-wavelength and high-order full-wavelength resonant modes. It can be noted that $\lambda_{0x}^{s}$ and $\lambda_{1x}^{l}$ are closer than $\lambda_{0y}^{l}$ and $\lambda_{1y}^{s}$. Therefore, the distortion of the magnitude curve of $T_{xx}$ is more significant than that of $T_{yy}$. The sharp phase shift in the phase curve at Fig.~\ref{result}(c) also implies the full-wavelength resonance. The periodicity of the unit cell is $7\times7$~mm$^2$, which is approximated to be $0.4 \lambda_0\times0.4 \lambda_0$ at $17.85$~GHz and the thickness is $0.048~\lambda_0$. The proposed design has a compact size compared to the previous designs at microwave frequencies (the unit-cell size is $0.48 \lambda_0\times0.48 \lambda_0\times0.15 \lambda_0$ at $11.8$~GHz in~\cite{motl}, and $0.4 \lambda_0\times0.4 \lambda_0\times0.28 \lambda_0$ at $10$~GHz in~\cite{AWPL}).

\section{Metasurface design}

\begin{figure}[h]
\centering
\includegraphics[width=0.9\columnwidth]{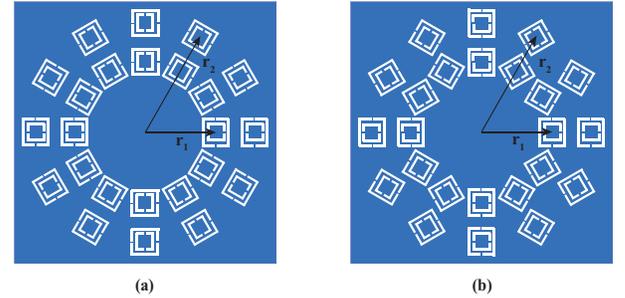}
\caption{Geometric structure of the metasurface. Topological charge is (a) $q=1$; (b) $q=2$. The radius of the inner ring is $r_1=14$~mm and that of the outer ring is $r_2=21$~mm.}
\label{whole}
\end{figure}

After optimizing the scatterers, a whole metasurface is built by using the scatterers with varying orientations as discussed in Section II. Two metasurfaces with topological charges of 1 and 2 are designed. The top view of the designed metasurface is shown in Fig.~\ref{whole}. Each metasurface includes $24$ scatterers, whose centers describe two circles with the radius of $r_1$ and $r_2$.

\subsection{Theoretical model}
The simulation domain for the metasurface with fine scatterers is large. Thus, the simulation process is very time-consuming. To facilitate fast and elegant design process, we propose a simplified theoretical model to predict the EM response of the metasurfaces.

\begin{figure}[h]
\centering
\includegraphics[width=0.9\columnwidth]{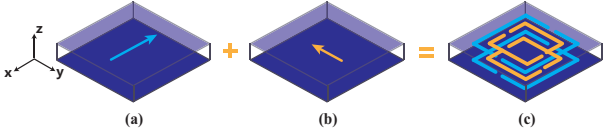}
\caption{Magnetic dipole modelling of the proposed scatterer.}
\label{gf}
\end{figure}

According to equivalent principle~\cite{chew}, the scattered field from CSRRs can be calculated by the equivalent magnetic currents distributed at the U-shaped slots. In view of subwavelength scale of unit cells, for simplicity, we represent the scatterer as two magnetic dipoles that are orthogonal to each other along $xy$ plane as shown in Fig.~\ref{gf}. The EM response of the whole metasurface is then equivalent to that from magnetic dipoles with varying orientations. Due to the different aperture sizes of the outer and inner CSRRs (See Fig.~\ref{final_unit}), the strengths of the two magnetic dipoles, which are proportional to the aperture areas, are different.

To start with, we write the $\mathbf{T}$ matrices for the outer and inner CSRRs as follows,
\begin{equation}
\mathbf{T}_{inner}
=\begin{pmatrix}
  0 & 0 \\ 0 & 1
\end{pmatrix},
\quad
\mathbf{T}_{outer}
=\begin{pmatrix}
  -1 & 0 \\ 0 & 0
\end{pmatrix}
\label{xy}
\end{equation}
The minus sign of the non-zero element in $\mathbf{T}_{outer}$ is introduced based on the designed $180^{\circ}$ phase difference between the co-transmission coefficients for the combined CSRRs.

Rotations of the CSRRs lead to new $\mathbf{T}$ matrices,
\begin{subequations}
\begin{align}
\mathbf{T}_{inner}(\alpha)
=\begin{pmatrix}
  \sin^2\alpha & -\sin\alpha\cos\alpha \\ -\sin\alpha\cos\alpha & \cos^2\alpha
\end{pmatrix}
\\
\mathbf{T}_{outer}(\alpha)
=\begin{pmatrix}
  -\cos^2\alpha & -\sin\alpha\cos\alpha \\ -\sin\alpha\cos\alpha & -\sin^2\alpha
\end{pmatrix}
\label{rxy}
\end{align}
\end{subequations}

At the incidence of circularly polarized wave, i.e., $\mathbf{E}_{i}=[1 ~-i]^T$, we can obtain the expression of the induced equivalent magnetic dipoles, having both $x$ and $y$ components,
\begin{subequations}
\begin{align}
\mathbf{M}_{inner}(\alpha)
=\mathbf{T}_{inner}(\alpha)\mathbf{E}_{i}
=\begin{pmatrix}
  M_{inner,x} \\  M_{inner,y}
\end{pmatrix}
=i e^{-i\alpha}
\begin{pmatrix}
   \sin \alpha  \\ -\cos\alpha
\end{pmatrix}
\\
\mathbf{M}_{outer}(\alpha)
=\mathbf{T}_{outer}(\alpha)\mathbf{E}_{i}
=\begin{pmatrix}
  M_{outer,x} \\  M_{outer,y}
\end{pmatrix}
=-e^{-i\alpha}
\begin{pmatrix}
   \cos\alpha  \\  \sin\alpha
\end{pmatrix}
\end{align}
\label{msource}
\end{subequations}

Then, the scattered electric field from the equivalent magnetic dipoles $\mathbf{M(r')}$ can be calculated by,
\begin{equation}
\begin{split}
\mathbf{E(r)}=2\int_V \overline{\mathbf{G}}_m(\mathbf{r,r'}) \cdot \mathbf{M(r')} d\mathbf{r'}\\
=2 \int_V
\bigtriangledown g(\mathbf{r,r'}) \times  \mathbf{M(r')} d\mathbf{r'}
\end{split}
\end{equation}
where $\overline{\mathbf{G}}_m(\mathbf{r,r'})$ is the magnetic dyadic Green's function, $g(\mathbf{r,r'})=\frac{e^{i k_0 R}}{4 \pi R}$ is the scalar Green's function, $R=|\mathbf{r-r'}|$ and $\mathbf{M(r')}$ is a point source expressed by Eq.~\ref{msource}.

By distributing all the magnetic dipoles corresponding to the location of the CSRRs in Fig.~\ref{whole}, the total radiated electric field can be obtained by summing up the electric fields scattered by the magnetic dipoles. This approach is useful to simulate a metasurface comprising a great number of scatterers that can be treated as point sources with arbitrary strengths, locations and orientations.

\begin{figure}[h]
\centering
\includegraphics[width=0.9\columnwidth]{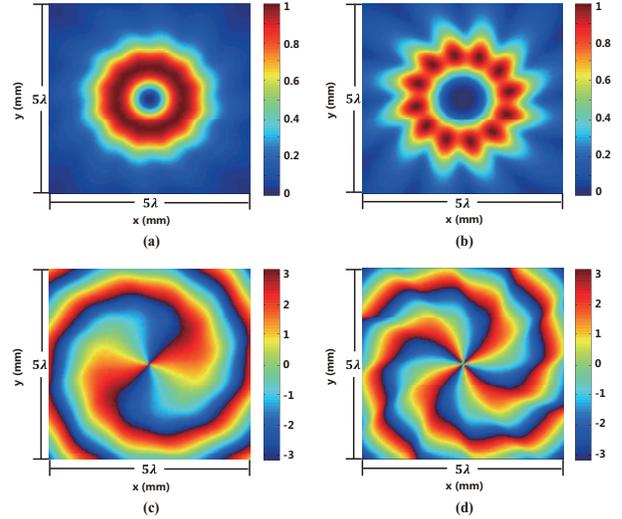}
\caption{Amplitude and phase distributions of the cross-circularly polarized component of electric field at a transverse plane of $z=10$~mm calculated by the proposed model. (a,b)  normalized amplitude; (c,d) phase in radian; (a,c) magnetic dipoles are distributed as Fig.~\ref{whole}(a); (b,d) magnetic dipoles are distributed as Fig.~\ref{whole}(b).}
\label{analytical}
\end{figure}

Using the proposed model, the patterns of electric field of the proposed metasurfaces are shown in Fig.~\ref{analytical}. The phase distribution suggests that the radiated EM wave has a helical wavefront carrying OAM. The phase changes $4\pi$ along a circular path around the phase singularity in Fig.~\ref{analytical}(c) and $8\pi$ in Fig.~\ref{analytical}(d). Therefore, the radiated EM wave carries OAM of order $2$ when $q=1$ and order $4$ when $q=2$.

\subsection{Simulation}

For full-wave simulation, a right-handed circularly polarized (RHCP) Gaussian beam is used to excite the two metasurfaces in Fig.~\ref{whole}. The reason why we use the Gaussian beam is to eliminate the truncation effect caused by the edges of the complementary surfaces. Specifically, the frequency of the Gaussian beam is set to be $17.85$~GHz; amplitudes of both the $x$ and $y$ components of electric field are $1$~V/m; the focal spot is located at the center of the whole metasurface and the beam waist is set to be $30$~mm (larger than the radius of the outer circle, $r_2=21$~mm in Fig.~\ref{whole}). Then the whole metasurface can be considered as being uniformly illuminated.

\begin{figure}[h]
\centering
\includegraphics[width=0.9\columnwidth]{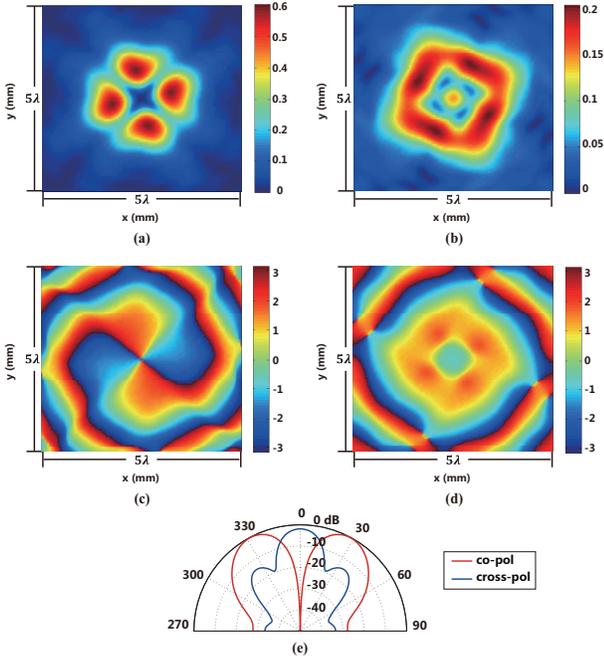}
\caption{Amplitude and phase distributions of the cross- and co-circularly polarized components of electric field at a transverse plane of $z=10$~mm calculated by the full-wave simulation and the forward far-field radiation pattern in $xz$ plane. Scatterers are distributed as Fig.~\ref{whole}(a). (a,b)  amplitude; (c,d) phase in radian; (a,c) cross-circularly polarized component; (b,d) co-circularly polarized component; (e) forward radiation pattern.}
\label{simulation1}
\end{figure}

\begin{figure}[h]
\centering
\includegraphics[width=0.9\columnwidth]{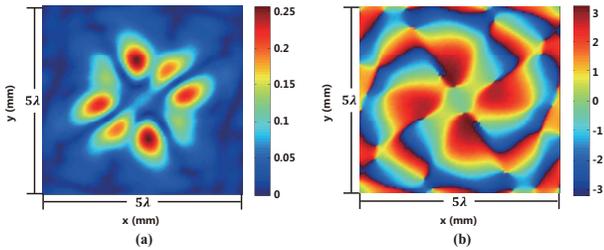}
\caption{Amplitude and phase distributions of the cross-circularly polarized component of electric field at a transverse plane of $z=10$~mm calculated by the full-wave simulation. Scatterers are distributed as Fig.~\ref{whole}(b). (a) amplitude; (b) phase in radian.}
\label{simulation2}
\end{figure}

For $q=1$, both the transmitted cross- and co-circularly polarized components of the electric field are shown in Fig.~\ref{simulation1}. As expected, the cross-circularly polarized component carries an OAM of order $2$ while the co-circularly polarized component carries no OAM. When $q=2$, the generated OAM order is $4$, which can be found in Fig.~\ref{simulation2}. The simulated cross-circularly polarized component shows similar phase pattern to its analytical pattern in Fig.~\ref{analytical}. Compared to the field pattern of OAM of order $2$, the pattern of OAM of order $4$ shows a more significant difference from that calculated based on the approximate magnetic dipole model. Firstly, the high-order OAM generation is more sensitive to the nonuniform Gaussian beam. Secondly, due to the periodicity breaking in the azimuthal direction as shown in Fig.~\ref{whole}(b), high-order diffraction from each unit cell becomes stronger.

\subsection{Efficiency}
It is known that the EM energy can be calculated by
\begin{equation}
U
=\oint_{A}\mathbf{S}\cdot d\mathbf{A}
\label{xy}
\end{equation}
where $A$ is a closed surface and $\mathbf{S}$ is the Poynting vector.

We first do a comprehensive efficiency analysis for our proposed metasurface of $q=1$. We choose $A$ as an infinitely large surface containing the observation plane of Fig.~\ref{simulation1}. If we consider the conversion efficiency as the ratio of the energy carried by the OAM wave to the total energy of the transmitted wave, the conversion efficiency is calculated to be $81.8\%$. We can also notice that the amplitude of the co-circularly polarized component is insignificant compared to that of the cross-circularly polarized one by checking Fig.~\ref{simulation1}(a) and (b). However, the ratio of the energy in the OAM wave to the incident Gaussian beam is $15.2\%$ and the ratio is predicted to be $81\%$ by the simulation results of the periodic unit cells in Section III B. This is because the periodicity along the $x$ and $y$ directions cannot be conserved and the truncation effect exists for the metasurface, where there is a limited number of scatterers. The forward far-field radiation pattern is shown in Fig.~\ref{simulation1}(e). We see a radiation null at the broadside and the main lobe is directed at $\theta=30^{\circ}$ for the cross-circularly polarized component, while at the same direction, for the co-circularly polarized component, the radiation is very weak.

\begin{figure}[h]
\centering
\includegraphics[width=0.9\columnwidth]{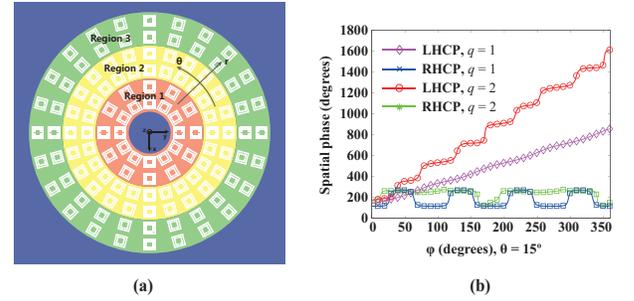}
\caption{(a) Schematic of the six-circle configuration when $q=1$; (b) far-field phase of both the co- and cross-circularly polarized components of electric field at a spherical angle of $\theta=15^\circ$ for $q=1$ and $q=2$.}
\label{whole_bigger}
\end{figure}

Next, we extend the two-circle metasurface in Fig.~\ref{whole}(a) to six-circle configuration. We divide the whole metasurface into three regions as shown in Fig.~\ref{whole_bigger}(a). Due to the truncation effect by the finite size, the periodicity along the $\theta$ and $r$ directions is only preserved in Region $2$. To save the computational cost, we illuminate this metasurface aperture with a RHCP plane wave. We then calculate the conversion efficiency defined by the energy of the transmitted wave on each region divided by the energy of the incident EM wave illuminated on the corresponding region at $z=\lambda_0$. The efficiencies are calculated to be $39\%$, $55\%$ and $42\%$ for Region $1$, $2$ and $3$ respectively. For the case when $q=2$, similarly, we calculate the three efficiencies which are $36\%$, $49\%$ and $40\%$. In both cases, the efficiency is the highest in Region $2$, where the periodicity is best preserved. Therefore, it is reasonable to conclude that by increasing the number of scatterers on the metasurface, the conversion efficiency can be improved. At a spherical angle of $\theta=15^\circ$, the far-field phase of both the co- and cross-circularly polarized components for $q=1$ and $q=2$ are drawn in Fig.~\ref{whole_bigger}(b). For $q=1$, the phase of the cross-circularly polarized component changes from $150^\circ$ to $870^\circ$ as $\phi$ increases from $0$ to $360^\circ$. The order of OAM is proved to be $2$ as expected. When $q=2$, the total phase change along the azimuthal direction is $4$ times $360^\circ$, indicating an OAM order $4$. The small fluctuations on the phase curve are caused by the high-order diffraction from unit cells.

\section{Conclusion}
In summary, to generate an OAM at microwave frequencies, we proposed an ultrathin complementary metasurface with the scatterers of sub-wavelength scale. The circuit model was established during the design process to reveal the working physics and facilitate the design optimization. Moreover, the whole metasurface was modelled by the equivalent magnetic dipoles and by applying dyadic Green's function, the deduced field patterns show a good agreement with the full-wave simulation. The efficiency of our proposed metasurface is carefully investigated. Thanks to the complementary design, a high transmission efficiency can be potentially achieved for arbitrary-order OAM generation by increasing the number of the scatterers.

\section{Acknowledgement}
This work was supported in part by the Research Grants Council of Hong Kong (GRF 716713, GRF 17207114, and GRF 17210815), NSFC 61271158, Hong Kong ITP/045/14LP, and Hong Kong UGC AoE/P¨C04/08.
\bibliographystyle{IEEEtran}
\bibliography{reference}

\end{document}